\documentclass[twocolumn]{aastex631}

\usepackage{amsmath}
\usepackage[normalem]{ulem}

\newcommand{\nuInu}{\ensuremath{
    \nu I_\nu
}}

\newcommand{\mmu}{\boldsymbol{\mu}}
\newcommand{\gilmore}{\mathrm{G12}}
\newcommand{\andrews}{\mathrm{A18}}
\newcommand{\crpropa}{\mathrm{\texttt{CRPropa}}}
\newcommand{\simprop}{\mathrm{\texttt{SimProp}}}

\begin{document}

\title{Impact of uncertainties on the cosmic optical and infrared backgrounds\\ on the propagation of astroparticles}

\author[0000-0003-0627-8436]{Lucas Gréaux}
\affiliation{Fakultät für Physik \& Astronomie, Ruhr-Universität Bochum, D-44780 Bochum, Germany}

\author[0000-0001-5681-0086]{Antonio Condorelli}
\affiliation{Université Paris Cité, CNRS, Astroparticule et Cosmologie, F-75013 Paris, France}

\author[0000-0003-1494-2624]{Leonel Morejon}
\author[0000-0002-2805-0195]{Karl-Heinz Kampert}
\affiliation{Bergische Universit\"{a}t Wuppertal, Gaußstraße 20, 42119 Wuppertal, Germany}

\author[0000-0002-4202-8939]{Jonathan Biteau}
\affiliation{Université Paris-Saclay, CNRS/IN2P3, IJCLab, 91405 Orsay, France}
\affiliation{Institut Universitaire de France (IUF)}

\begin{abstract}

When propagating through the universe, gamma rays at very-high energy (VHE, $E > 100$\,GeV) and ultra-high energy cosmic rays (UHECRs, $E>1$\,EeV) can interact with the optical, infrared and microwave photon fields that permeate the universe. These interactions result in a characteristic absorption imprint in the spectra of extragalactic gamma-ray sources at VHE, and in a change in the mass-composition of UHECRs. The study of both VHE gamma rays and UHECRs therefore requires precise knowledge of the intensity of the cosmic photon fields. In this work, we explore the impact of the current uncertainties on the optical and infrared photon fields on the propagation of astroparticles. We restrict the range of available models to those best matching the recent measurements, and compare the different reconstructions resulting from these models. We find that the knowledge on the cosmic background light is no longer the dominant source of uncertainties in understanding the phenomenology of astroparticle sources in the low redshift universe $(z<0.1)$, enabling robust spectral and composition inference with the next generation of both gamma-ray and UHECR measurements.

\end{abstract}


\section{Introduction} \label{sec:intro}

The electromagnetic radiation emitted throughout the history of the universe following the emission of the cosmic microwave background (CMB)  covers about 20 decades in wavelength \citep[see e.g.][]{Cooray_2016, 2021arXiv210212089D, Biteau_2025}, including the cosmic radio background (CRB; 10\,MHz -- 10\,GHz), cosmic infrared background (CIB; 8\,\textmu m -- 2\,mm), cosmic optical background (COB; 0.1 -- 8\,\textmu m), cosmic X-ray background (CXB; 0.3\,keV -- 2\,MeV), and cosmic gamma-ray background (CGB; 2\,MeV -- 1\,TeV). These fields are collectively referred to as extragalactic background light (EBL), though this term is occasionally used to refer specifically to the CIB and COB \citep[see e.g.\ review by][]{Dwek:2012nb}.

The EBL plays an important role in the propagation of astroparticles. Gamma rays with very-high energies (VHE; $E > 100$\,GeV) can interact with optical and infrared EBL photons, and at higher energies ($E > 100$\,TeV) with photons from the CMB. This results in the production of electron-positron pairs ($\gamma + \gamma \rightarrow e^+ + e^-$) through the Breit–Wheeler process \citep{Gould_Schreder_1967a, Gould_Schreder_1967b}. While these pairs can in turn produce lower energy gamma rays by upscattering CMB photons, deflections by the intergalactic magnetic fields are expected to spread this emission.  The secondary gamma rays may effectively be considered lost, although the search for their dim signature remains an active field of research \citep[see e.g. reviews by][]{2021Univ....7..223A, 2022Galax..10...39B}. Typically, the mean free path (proper attenuation length) of gamma rays at $1$\,PeV,  $10$\,TeV, and $100$\,GeV is on the order of 10\,kpc, 100\,Mpc, and 30\,Gpc, respectively. 

Ultra-high energy cosmic rays (UHECRs; $E > 1$\,EeV) are composed of protons and nuclei and can also interact with electromagnetic backgrounds \citep[see e.g.][]{Berezinsky:2005cq,Allard_2006}. Protons can undergo pair production ($p + \gamma \rightarrow p + e^+ + e^-$) and pion production ($p + \gamma \rightarrow p + \pi^0$ and $p + \gamma \rightarrow n + \pi^+$). Heavier nuclei suffer single- and multiple-nucleon photodisintegration (e.g.\ $A \to (A-1) + p,n$) mostly with the CMB and CIB. Through these interactions, heavy elements can become lighter as they lose energy, resulting in a change in the composition of the UHECR flux. 

The main contributor to the EBL is the integrated galaxy light (IGL), the combined emission from galaxies. The advent of deep-field spaceborne instruments, such as the Hubble Space Telescope (HST), and ground-based instruments, such as the Very-Large Telescope (VLT), has enabled robust measurements of the number of galaxies at different wavelengths. These measurements have been used to derive the IGL with 2-20\,\% precision between 0.1 and 500\,\textmu{m} \citep{Driver:2016ApJ...827..108D, Koushan:2021MNRAS.503.2033K, Tompkins:2026MNRAS.547ag044T}. Additional unresolved contributions to the EBL can still be expected, such as diffuse stellar halos around galaxies and light remnants from sources that reionized the universe \citep{Cooray_2012}. These emissions are not included in the IGL, but they can be accounted for by measuring the surface brightness of the night sky directly, albeit with a lower precision than IGL measurements at comparable wavelengths \citep{Mattila_2017, Postman:2024ApJ...972...95P}. Such direct measurements are now  free of contamination from orders of magnitude brighter foregrounds, particularly from zodiacal light  \citep[sunlight refracted by interplanetary dust, see e.g.][]{Korngut_2022, Biteau_2025}. This is especially the case for measurements taken outside the inner solar system \citep{Postman:2024ApJ...972...95P}. Other measurements of the EBL can account for diffuse components in addition to the IGL. Cross-correlations of observations in different bands allow for EBL measurements that are free from foreground contamination \citep{Chiang:2019ApJ...877..150C}. Additionally, the interaction between gamma rays and the EBL enables comprehensive measurements of the amount of light along the line of sight, which show remarkable agreement with the IGL \citep[][see Sec.~\ref{sec:gamma}]{Biteau_Williams_2015, Greaux_2024}.

Although earlier studies have quantified the impact of different EBL models on the simulated spectrum and composition of UHECRs on Earth \citep[see e.g.][]{AlvesBatista:2015jem}, the precision of these models has since improved substantially. In this study, we examine the impact of state-of-the-art EBL models on the propagation of VHE gamma rays and UHECRs. This paper is organized as follows:
In Sec.\,\ref{sec:comparison}, we test a set of EBL models against wavelength-resolved observations of the extragalactic sky's intensity to select the models that best match the current knowledge of the EBL.
In Sec.\,\ref{sec:gamma}, we demonstrate how these models impact the propagation of gamma rays using observations from the current generation of imaging atmospheric Cherenkov telescopes.
In Sec.\,\ref{sec:cosmic-rays}, we study how these models affect the propagation of UHECRs.
Finally, in Sec.\,\ref{sec:conclusions}, we conclude on the overall impact of the current uncertainties on the optical and infrared backgrounds on the propagation of astroparticles.

\section{Comparison of the existing models with measurements of the EBL;} \label{sec:comparison}

\subsection{Different EBL models}
\label{sec:EBL_models}

\begin{figure}[t]
    \centering
    \includegraphics[width=\columnwidth]{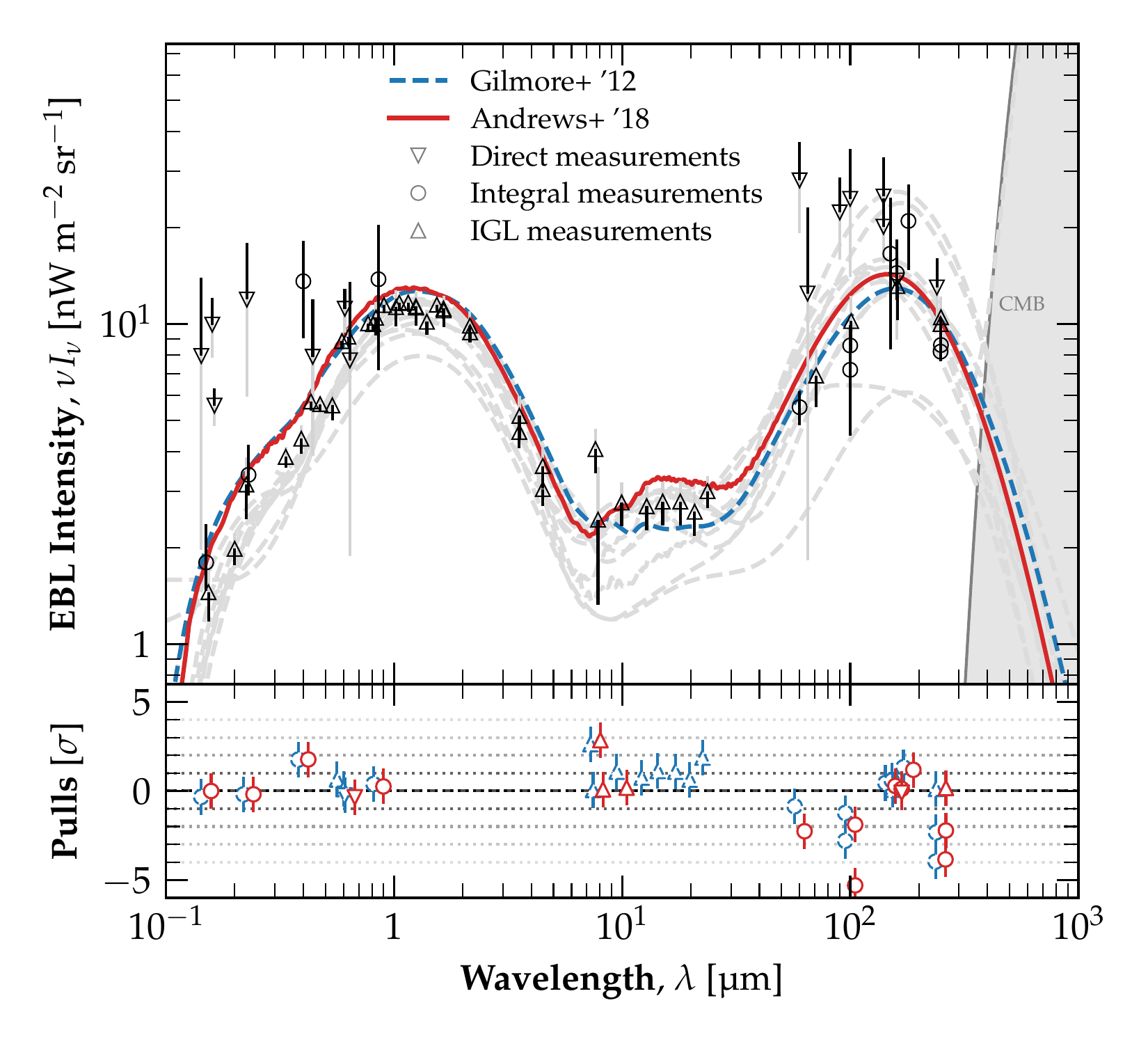}
    \caption{
        \textit{Top:}
        Intensity of the COB and the CIB as a function of wavelength $\lambda$.
        Measurements of the IGL are shown with upward-pointing triangles \citep{Milliard:1992A&A...257...24M, Driver:2016ApJ...827..108D, Duivenvoorden:2020MNRAS.491.1355D, Koushan:2021MNRAS.503.2033K, Windhorst:2023AJ....165...13W, Kim:2024MNRAS.527.5525K, Tompkins:2026MNRAS.547ag044T}, direct EBL measurements with downward-pointing triangles \citep{Murthy:1989ApJ...336..954M, Murthy:1990A&A...231..187M, Martin:1991ApJ...379..549M, Brown:2000AJ....120.1153B, Finkbeiner:2000ApJ...544...81F, Odegard:2007ApJ...667...11O, Matsuoka:2011ApJ...736..119M, Matsuura:2011ApJ...737....2M, Postman:2024ApJ...972...95P}, and integral measurements with circles \citep{Juvela:2009A&A...500..763J, Marsden:2009ApJ...707.1729M, Penin:2012A&A...543A.123P, Mattila:2017MNRAS.470.2152M, Chiang:2019ApJ...877..150C, Chiang:2025ApJ...992...65C, Haikala:2026A&A...706A.358H}. Measurements contaminated by zodiacal light have been omitted \citep[see][]{Biteau_2025}.
        The selected models, with the smallest deviance with respect to the data, are shown in dashed blue \citep{Gilmore_2012}, and solid red \citep{Andrews_2018}. Other models that are in tension with observations are shown in dashed gray \citep{Franceschini_Rodighiero_2017, Baes_2019, Lagos_2019, Khaire_Srianand_2019, SaldanaLopez_2021, Koushan:2021MNRAS.503.2033K, Finke_2022, PorrasBedmar_2025}. For a quantitative comparison of deviances, see Table \ref{tab:ebl_deviance}.
        \textit{Bottom:} Normalized difference between the data and the selected EBL models. The points have been offset along the x-axis for clarity. 
        The code and data are available in \citet{biteau_jonathan_2023_7842239}.
    } \label{fig:EBL_models_densities}
\end{figure}

Figure\,\ref{fig:EBL_models_densities} shows the different EBL models considered in this work. These models are divided into three main classes, according to the classification of 
\citet{Pueschel_Biteau_2021}.

The first category encompasses so-called empirical models \citep[e.g.][]{Dominguez_2011, Franceschini_Rodighiero_2017, SaldanaLopez_2021}. These models are based on galaxy number counts as a function of redshift, interpolated through the redshift-dependent luminosity functions of galaxies. In the most recent works, the EBL is reconstructed from the evolution of galaxy population emissions up to $z = 6$ \citep[see e.g.][]{SaldanaLopez_2021}.

Phenomenological models, such as those presented in \citet{Driver:2016ApJ...827..108D, Andrews_2018, Finke_2022, PorrasBedmar_2025}, are based on the emissions of star populations, the birth distribution of which is determined from a universal initial mass function. A stellar population synthesis model is then used to derive the cumulative spectrum of these stars, weighted by the cosmic star formation rate and integrated throughout cosmological epochs.
The fraction of photons escaping absorption by dust as a function of redshift is parameterized using templates or models of broadband galaxy spectra, with the remaining fraction reradiated in the CIB.

\textit{A priori} models \citep[e.g.][]{Gilmore_2012, Lagos_2019} are the third group of models considered. These models start from $N$-body simulations of dark matter halo mergers. Sub-grid baryonic physics recipes are added to simulate star formation and black hole accretion. Finally, components similar to those in the phenomenological models are incorporated to reproduce the currently observed COB and CIB. 

We do not include in this study the most recent models from \citet{Driver_2026}, as predicting their impact on astroparticle sources would require the production of propagation tensors not yet evaluated by the community. Nonetheless, the models that we select encompass the resolution on the EBL spectrum as estimated by \citet{Driver_2026}.

\subsection{Comparison to EBL measurements}
\label{sec:comparison_models}

The three types of models considered -- empirical, phenomenological, and \textit{a priori} -- have been calibrated by their authors using observational data. Since the first detection of the CIB \citep{Puget_1996}, deep field surveys have enabled the counting of galaxies at sufficiently high magnitudes to reach uncertainties on the intensity of the IGL down to few percent. At optical wavelengths, cosmic variance now dominates the error budget, highlighting the importance of deep, full-sky surveys \citep{Tompkins:2026MNRAS.547ag044T}. Following these improvements, the level of detail and accuracy of EBL models has dramatically evolved over the past two decades.

We compare the different EBL models introduced in Sec.~\ref{sec:EBL_models} to the compilation of EBL measurements ranging from 0.1 to 300\,\textmu{m} collected by \cite{Driver_2026}. Following the authors’ suggestion, we include an additional $5\%$ uncertainty added in quadrature to each IGL data point, to account for the minimal expected cosmic variance in the different surveys. Above 300\,\textmu{m}, the intensity of the CMB overwhelms that of the EBL, yielding no measurable effect on the propagation of astroparticles. For each model, $\nuInu^m$, we compute a deviance $D_m$ with respect to the 75 selected data points. Considering the Heaviside step function $\Theta$, the deviance is defined as $D_m = D_\textrm{LL} + D_\textrm{UL} + D_\textrm{M}$, with
\begin{align} \label{eq:ebl_deviance_breakdown}
    D_\textrm{LL} & = \sum_{i \in \textrm{LL}}
    \Theta\left( \nuInu^m(\lambda_i) - \nuInu^i \right)
    \left( \frac{\nuInu^m(\lambda_i) - \nuInu^i}{\sigma_i}\right)^2 \textrm{,}  \notag \\
    D_\textrm{UL} & = \sum_{i \in \textrm{UL}}
    \Theta\left( \nuInu^i - \nuInu^m(\lambda_i) \right)
    \left( \frac{\nuInu^m(\lambda_i) - \nuInu^i}{\sigma_i}\right)^2 \textrm{,} \notag \\
    D_\textrm{M} & = \sum_{i \in \textrm{M}}
    \left( \frac{\nuInu^m(\lambda_i) - \nuInu^i}{\sigma_i}\right)^2 \textrm{,}
\end{align}
where $\lambda_i$, $\nuInu^i$, and $\sigma_i$ are the wavelengths, intensities, and uncertainties on the intensities, respectively. In this comparison, IGL measurements are treated as lower limits (LL) contributing to the deviance only when they are higher than the model value. Conversely, direct measurements are treated as upper limits (UL) and contribute only when they are lower than the model value.

\begin{table}[!t]
    \centering
\begin{tabular}{lrr}
\hline \hline
EBL model & $D_\mathrm{EBL}$ & eDoF \\
\hline
G12 \citep{Gilmore_2012} & 52.6 & 25 \\
A18 \citep{Andrews_2018} & 69.8 & 18 \\
\citet{PorrasBedmar_2025} (Chary) & 146.4 & 38 \\
\citet{Finke_2022} & 178.4 & 45 \\
\citet{Khaire_Srianand_2019} & 184.5 & 50 \\
\citet{PorrasBedmar_2025} (2BB) & 187.5 & 41 \\
\citet{PorrasBedmar_2025} (BOSA) & 189.7 & 39 \\
\citet{Franceschini_Rodighiero_2017} & 196.0 & 40 \\
\citet{SaldanaLopez_2021} & 255.1 & 41 \\
\citet{Lagos_2019} & 636.3 & 42 \\
\citet{Franceschini_2008} & 770.0 & 34 \\
\citet{Dominguez_2011} & 1167.3 & 39 \\
\citet{Baes_2019} & 1216.8 & 44 \\
\hline
\end{tabular}
    \vspace{10pt}
    \caption{
        Deviance $D_\textrm{EBL}$ and estimated effective number of degrees of freedom (eDoF) for each EBL model, with respect to EBL measurements between 0.1 and 300\,\textmu{m}.
    }\label{tab:ebl_deviance}
\end{table}

The deviances computed for each EBL model are reported in Table~\ref{tab:ebl_deviance}. For most models, the deviance is much higher than the effective number of degrees of freedom, estimated here as the number of points with a non-zero contribution in Eq.~\eqref{eq:ebl_deviance_breakdown}. Such high reduced deviances are expected since the data corpus includes more recent measurements than were used to develop most of the models. In particular, the recent precise measurements from \citet{Chiang:2025ApJ...992...65C} and \citet{Tompkins:2026MNRAS.547ag044T} have the greatest impact on the deviances.

From the sample of EBL models, two models stand out as having both low computed deviance $D_\textrm{EBL}$ and low effective number of degrees of freedom, eDoF. These models, the \textit{a priori} model from \citet{Gilmore_2012} and the phenomenological model from \citet{Andrews_2018}, are the ones showing the lowest tension with the measurements of the EBL at $z=0$. 
Although these models are consistent at $z=0$, they are based on different modeling approaches, which affect their evolution with redshift. We select these models to evaluate the impact of EBL uncertainties on the propagation of VHE gamma rays and UHECRs, and refer to them in the following as G12 and A18, respectively.
We have checked that applying scaling factors of $0.83$ and $0.88$ to G12 and A18, respectively, as provided by \citet{Driver:2016ApJ...827..108D} to match the IGL data, does not affect our conclusions.\footnote{It should be noted that \citet{Driver_2026} assume that the IGL data represent all of the EBL, leading to rescaling factors below unity, while we assume here that the IGL data only provide a lower limit on the EBL.}

\section{Impact of the selected EBL models on
gamma-ray propagation and comparison to data.} \label{sec:gamma}

\subsection{Gamma-ray data analysis}

The interaction between EBL photons and VHE gamma rays was first predicted in the 1960s \citep{Nikishov_1961, Gould_Schreder_1967a, Gould_Schreder_1967b}, three decades before the first constraints on the opacity of the Universe to gamma rays \citep{Stecker_1992}. In the following years, significant progress was made in modeling the EBL \citep{Franceschini_2008, Dominguez_2011, Gilmore_2012}, leading to the first detection of its absorption imprint in both the HE \citep{FERMI_2012} and VHE \citep{HESS_2013} regimes. Combining data from multiple IACTs, \citet{Biteau_Williams_2015} demonstrated that a model-independent measurement of the EBL at $z=0$ using gamma rays was feasible. More recently, \citet{Greaux_2024} showed the first agreement between direct, integrated galaxy light, and gamma-ray measurements of the EBL, in the first gamma-ray study including the redshift evolution as a free parameter.

The transparency of the EBL to gamma rays with energy $E$ emitted at redshift $z$ is characterized by the EBL optical depth, $\tau(E, z)$. An observer on Earth cannot access the intrinsic spectra of gamma-ray sources, $\phi_\mathrm{int}(E, z)$, but can only measure their observed spectra $\phi_\mathrm{obs}(E, z) = \phi_\mathrm{int}(E, z) \times e^{-\tau(E, z)}$. In this study, we examine how different EBL models, selected in Sec.~\ref{sec:comparison}, affect the observed gamma-ray spectra of extragalactic sources. We use the STeVECat catalog \citep{Greaux_2023}, from which we select spectra with at least four data points, excluding upper limits. Furthermore, we only consider sources with a known redshift $z > 0.01$, for which some EBL absorption is expected. This selection yields 268 spectra from 45 sources, and corresponds to the VHE data corpus used in \citet{Greaux_2024}. This dataset consists of a total of 2211 flux points, between 44 GeV and 25 TeV.

Since the intrinsic spectra of gamma-ray sources are unknown, they must be modeled. Due to the non-thermal nature of the emission, the chosen models are functions derived from the power law \citep[see][]{Biasuzzi_2019}: the power law (PWL), the log parabola (LP), the power law with exponential cutoff (EPWL), and the log parabola with exponential cutoff (ELP). For each spectrum and EBL model, we  iteratively select the best spectral template from these four models, as done in \citet{Biasuzzi_2019}. Starting with the PWL, a more complex model is chosen if it is favored by the data at the $2\sigma$ level, considering the deviance $\chi_\gamma^2$ defined as
\begin{equation}
    \chi_\gamma^2 = \sum_{E_i, \phi_i, \sigma_i} \left(
    \frac{\phi_m(E_i) - \phi_i}{\sigma_i}
    \right)^2\textrm{,}
\end{equation}
where $E_i$, $\phi_i$, and $\sigma_i$ are the energies, fluxes, and uncertainties of the VHE spectrum, and $\phi_m$ is the model under consideration.

We apply this iterative procedure to the corpus of gamma-ray spectra, using the EBL models G12 and A18. We show in Figure\,\ref{fig:gamma_spectrum} the best-fit spectra obtained for an observation of the blazar Markarian 501, located at a redshift $z=0.033$ \citep{MAGIC_2020}. For both EBL models, the preferred spectral shape is the EPWL. The computed deviances are $\chi^2_\gilmore = 9.0$ and $\chi^2_\andrews = 9.4$, with 14 degrees of freedom. The data is well represented by the model, and the reconstructed $p$-values are similar for G12 and A18. The residuals, or pulls, defined as the difference between the data and the best-fit model normalized to the uncertainty of the data, are compatible with zero across the energy range, indicating a good match between the reconstructed spectra and the data.

\subsection{Impact on the reconstruction of observed spectra}

\begin{figure}[!t]
    \centering
    \includegraphics[width=\columnwidth]{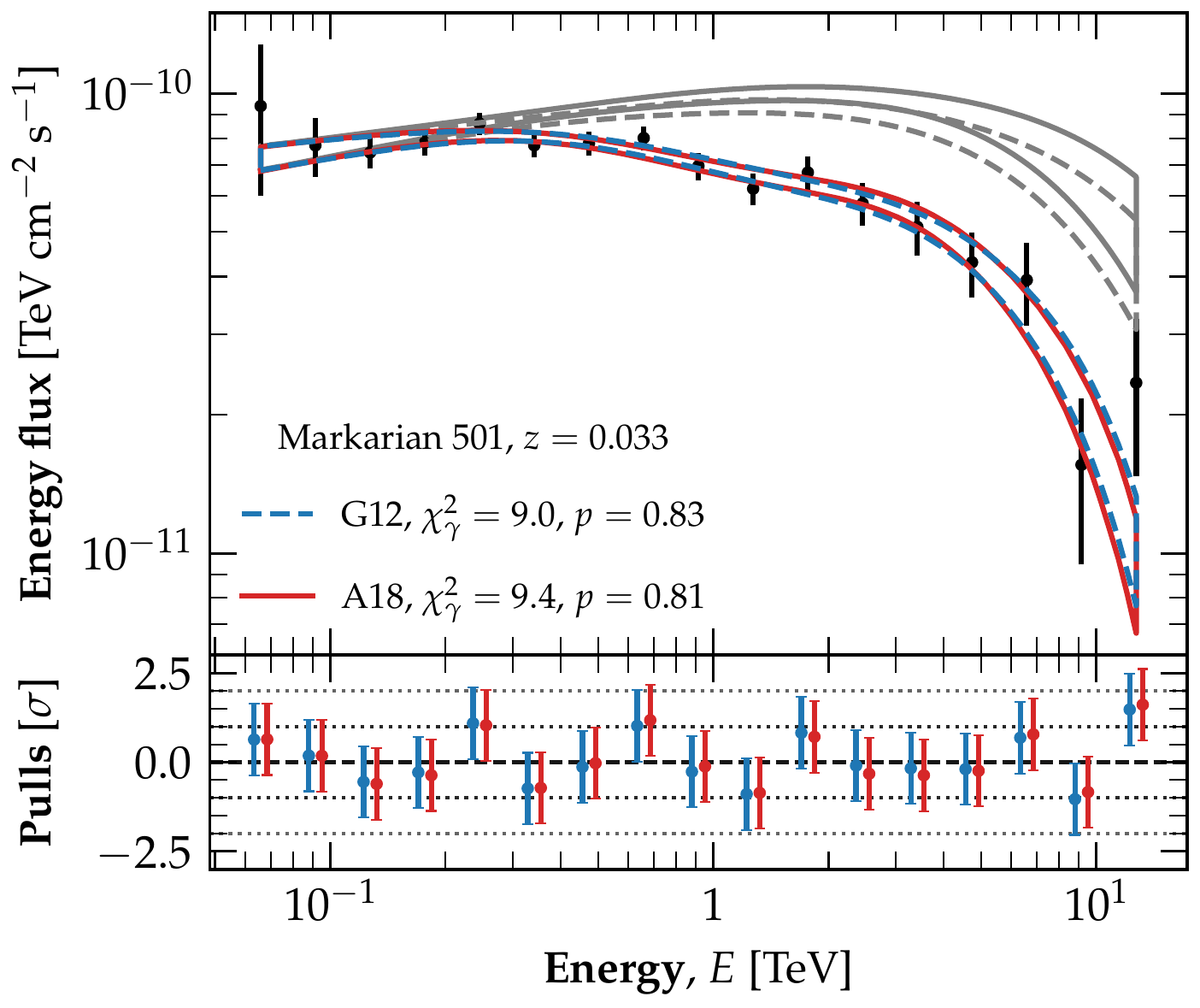}
    \caption{
        \textit{Top}: Best-fit spectra for a MAGIC observation of the blazar Markarian 501 \citep{MAGIC_2020}, obtained using the two selected EBL models. Each reconstructed spectrum at Earth is shown as a colored bow-tie, corresponding to the $1\sigma$ uncertainty band. The corresponding spectra without EBL absorption are shown in gray.
        \textit{Bottom}: Normalized difference between the gamma-ray data and the best-fit spectra. The points have been offset along the x-axis for clarity.
    } \label{fig:gamma_spectrum}
\end{figure}

The behavior shown in Figure\,\ref{fig:gamma_spectrum} is consistent across the entire gamma-ray dataset. A total of 196 spectra are modeled for G12 using the PWL, 35 using the LP, 32 using the EPWL, and five using the ELP. In total, this corresponds to 1,598 degrees of freedom. For A18, the preferred models are 197 PWL, 39 LP, 27 EPWL and 5 ELP, corresponding to 1,599 degrees of freedom. For these two EBL models, the overall deviances are $\chi^2_\gilmore = 1450$ and $\chi^2_\andrews = 1469$. These values are smaller than the number of degrees of freedom, showing that the gamma-ray data tend to be overfitted. This is expected, as correlations between data points are not reported with the spectra and cannot therefore readily be accounted for in an analysis of archival data.

In order to directly compare the two selected EBL models, we consider a new, joint set of spectral templates. We create this new set by selecting for each gamma-ray spectrum the simplest template containing both the G12- and the A18-preferred templates. This joint set consists of 195 PWL, 35 LP, 28 EPWL and 10 ELP, which corresponds to 1,592 degrees of freedom.

\begin{figure}[!t]
    \centering
    \includegraphics[width=\columnwidth]{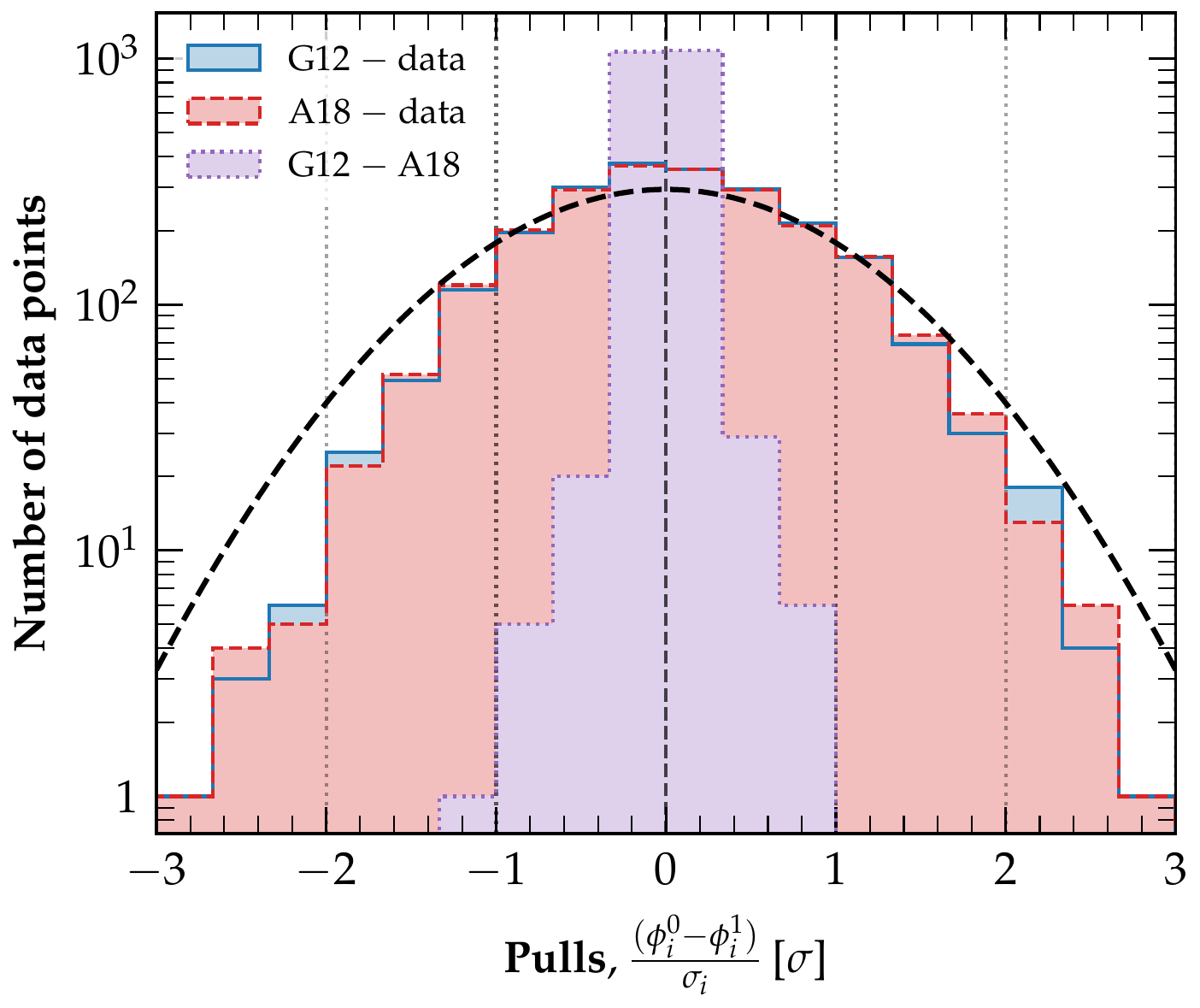}
    \caption{
        Distribution of the pulls of all the best-fit spectra across the entire data corpus, using the joint spectral templates.
        The pulls for the EBL models G12 and A18 are shown in dashed blue and solid red, respectively.
        The normalized differences between the best-fit spectra using the two EBL models are shown in dotted purple.
        For reference, the black dashed curve represents a standard normal distribution, scaled to the size of the dataset.
    } \label{fig:gamma_pulls}
\end{figure}

For each gamma-ray spectrum in the dataset, we computed the best-fit spectra using the joint spectral templates. We show in Figure\,\ref{fig:gamma_pulls} the distribution of pulls across the entire data corpus. For both G12 and A18, the pulls follow a Gaussian distribution centered around 0, with a standard deviation of approximately 1. Overall, the spectra are well modeled when using the two EBL models. We also show in Figure\,\ref{fig:gamma_pulls} the normalized difference between the best-fit spectra from the two models. This distribution is also centered around 0, but with a standard deviation significantly smaller than 1. The differences in reconstruction between the two models are negligible when compared to the uncertainties in the data.

To compare the overall preference of the gamma-ray data, we create a compound EBL model, the intensity of which is a linear combination of the intensities from G12 and A18. For a weight $0 \leq \mu \leq 1$, we define the composite model $\nuInu^\mu = \nuInu^\gilmore \times (1 - \mu) + \nuInu^\andrews \times \mu$, such that $\mu = 0$ and $\mu = 1$ correspond to the G12 and A18 models, respectively. Since the optical depth of an EBL model is directly proportional to its specific intensity, it follows that $\tau_\mu = \tau_\gilmore \times (1 - \mu) + \tau_\andrews \times \mu$.

We show in Figure\,\ref{fig:gamma_ebl_fit} the $\chi^2_\gamma$ profile obtained by summing the best-fit deviances from the complete gamma-ray dataset for different values of the weight $\mu$, using the joint set of spectral templates. As expected, the deviances are lower than those obtained using the model-preferred spectral templates, as the joint set allows for less degrees of freedom. We find that G12 is favored by the data compared to A18 at the $4.1\sigma$ level. The preferred parameter value, $\mu_\textrm{best} = -0.03 \pm 0.25$, is compatible with G12.

\subsection{Impact on gamma-ray parameter reconstruction}

As shown in Figure\,\ref{fig:gamma_pulls}, the individual differences between the best-fit observed spectra reconstructed using the EBL models G12 and A18 are negligible. However, the reconstruction of the intrinsic spectral parameter, free from EBL absorption, is expected to reflect the differences in transparency between the EBL models. For each gamma-ray spectrum, we compute the spectral index $\Gamma$ and the energy flux at the middle (in log-scale) of the reported energy-range.

\begin{figure}[!t]
    \centering
    \includegraphics[width=\columnwidth]{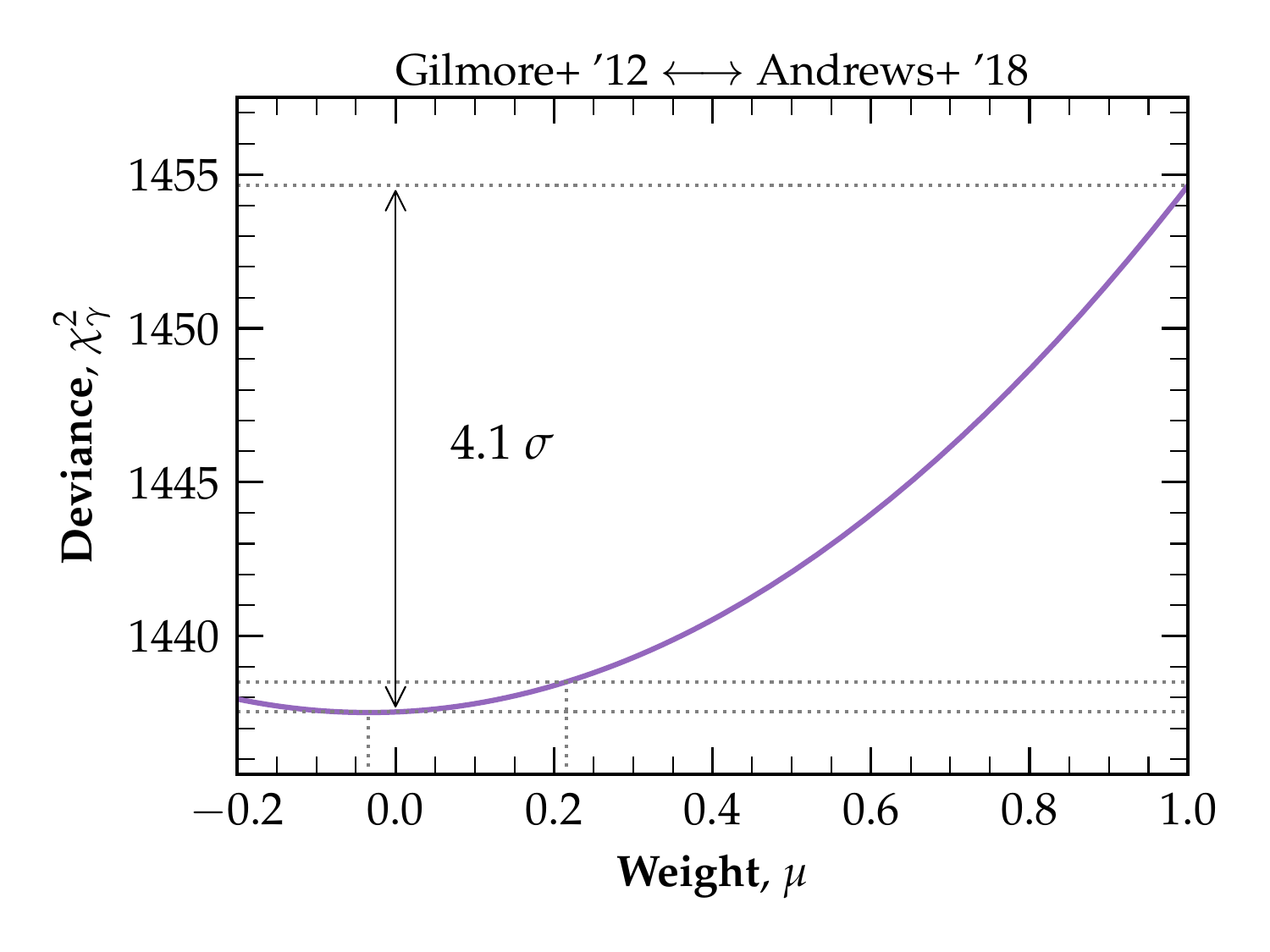}
    \caption{
        Deviance profile reconstructed for the constructed compound EBL model, defined from models $\nuInu^\gilmore$ and $\nuInu^\andrews$ as $\nuInu^\mu = \nuInu^\gilmore \times (1 - \mu) + \nuInu^\andrews \times \mu$.
        The preferred parameter value is $\mu_\textrm{best} = -0.03 \pm 0.25$.
    } \label{fig:gamma_ebl_fit}
\end{figure}

\begin{figure}[!t]
    \centering
    \includegraphics[width=\columnwidth]{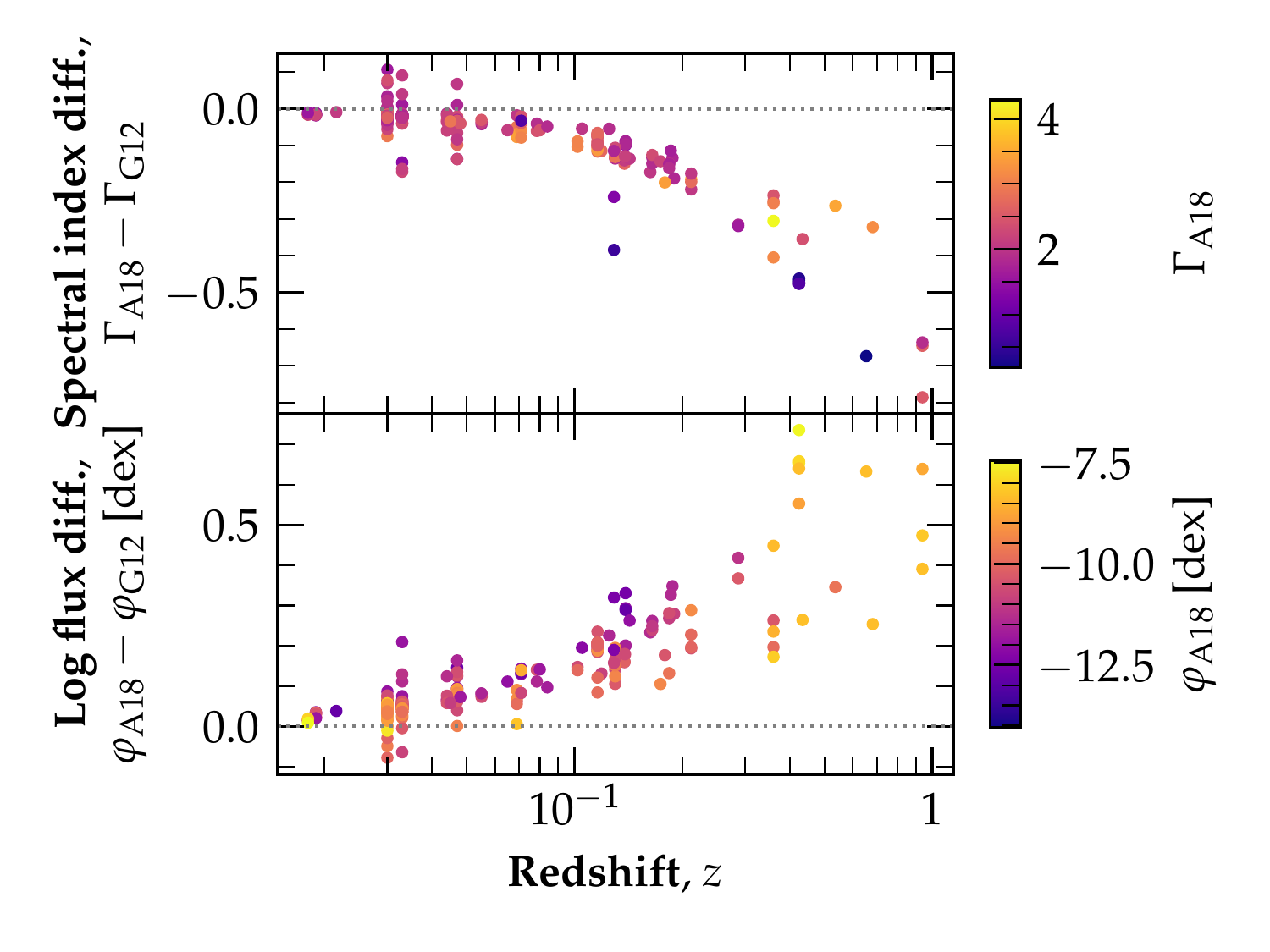}
    \caption{
        \textit{Top}: Difference in intrinsic spectral indices reconstructed using A18 and G12, as a function of source redshift.
        \textit{Bottom}: Difference in base-10 logarithm of intrinsic flux (in TeV\,cm$^{-2}$\,s$^{-1}$) reconstructed using A18 and G12.
        The color of the points correspond to the spectral index and base-10 logarithm of intrinsic flux, computed using A18.
    }\label{fig:gamma_indices}
\end{figure}

We show in Figure\,\ref{fig:gamma_indices} the difference between the intrinsic spectral indices reconstructed over the entire gamma-ray dataset, using the EBL models G12 and A18. We also show in Figure\,\ref{fig:gamma_indices} the differences between the base-10 logarithm of the intrinsic energy fluxes, $\varphi = \log_{10} \frac{\phi}{\phi_0}$ with $\phi_0 = 1\,$\,TeV\,cm$^{-2}$\,s$^{-1}$. The differences in spectral indices and energy fluxes appear to be anti-correlated, which is expected as the same data is well reproduced by the two models.

For sources below redshift 0.1, the fluxes and the spectral indices are different by less than 0.2\,dex and 0.2 units, respectively. This result is consistent with the agreement between the intensities of G12 and A18 at redshift $z=0$ shown in Figure\,\ref{fig:EBL_models_densities}. The differences in fluxes and indices become more pronounced at redshifts greater than 0.1. Sources at these redshifts provide the bulk of the evidence for rejecting A18 in favour of G12, as shown in Figure\,\ref{fig:gamma_ebl_fit}. These results demonstrate the importance of high-redshift gamma-ray spectra for distinguishing between different EBL evolution scenarios.

\section{Impact of the selected EBL models on UHECR propagation and comparison to data.} \label{sec:cosmic-rays}

\subsection{UHECR data analysis}

UHECR in the EeV energy range consist of a mixture of protons, helium, with an increasingly large proportion of ionized metals as the energy increases \citep{PierreAuger:2016qzj,PierreAuger:2024flk}. The observation of a large-scale, dipole-like anisotropies at energies greater than 8\,EeV has led to the conclusion that they are of extragalactic origin \citep{PierreAuger:2017pzq}.

Shortly after the discovery of the CMB and well before high-statistics measurements of UHECRs were available, it was noted that protons with energies greater than about 60\,EeV would lose energy through photo-pion production in the CMB and that nuclei would rapidly photodisintegrate. This interaction, known as the GZK-cutoff \citep{Greisen:1966jv, Zatsepin:1966jv}, limits their horizon to some tens to hundreds of Mpc, depending on energy and nuclear mass of UHECRs. The first in-depth study of photodisintegration processes in the CMB, which also addresses energy losses of protons through $e^+e^-$-pair-production, was presented by
\cite{Stecker1969}. Later, \cite{Aloisio:2006wv} pointed out that $p\to p\,e^+e^-$-pair-production in the CMB could naturally explain the so-called ``ankle'' at about 5\,EeV, assuming that the UHECR composition at this energy would be dominated by protons with a helium admixture of not more than about 15\,\%. However, such a light composition at the ankle has been excluded by observations \citep{PierreAuger:2016qzj}. 

The interaction processes previously described can also occur with the different components of the EBL, at different effective energies. The most important component in the context of this work is the CIB, which dominates the photo-pion and photodisintegration processes at Lorentz factors of $\log_{10}\Gamma \lesssim 9.5$ \citep[see e.g.][]{Allard:2011aa}. Such Lorentz factors correspond to $E/A \lesssim 3$\,EeV, i.e.\ to
energies of protons below the ankle in the UHECR spectrum and to energies of heavy nuclei up to the flux suppression. Previous studies of the effects of energy losses and changes of the chemical composition of UHECR due to their propagation through the EBL can be found in 
\citet{Khan:2004nd,Allard:2011aa,Kampert:2012fi,Unger_2015PhRvD..92l3001U}. These works highlight the importance of a proper modeling of UHECR propagation in order to infer, from the observations made at Earth, the UHECR energy spectra and mass composition emitted by astrophysical sources.

In this study, we compare the simulated extragalactic propagation of UHECRs to observed data from the Pierre Auger Observatory, using public software developed in the context of the MICRO project.\footnote{Multi-messenger Research on Cosmic Ray Origins (MICRO), \url{https://micro-uhecr.github.io}} We use the same framework as the one developed in previous works by \citet{Luce_2022, Marafico_2024}. The goal of these studies was to define an astrophysical scenario assuming bursting sources that evolve according to  the cosmic star formation rate density. For each nuclear mass $j$ at the source, the injected flux is written as:
\begin{equation}
    J_j (E_{i}) = \varepsilon^{\rm tot}_j \cdot k \cdot \bigg(\dfrac{E_{i}}{E_{0}} \bigg)^{-\gamma} \cdot f_{\rm{cut}} (E_{i}, Z \cdot R_{\rm{cut}}) \textrm{,}
    \label{eq:inj_spectrum}
\end{equation}
where $E_{0}$ is a reference energy, $E_i$ is the energy of a given bin in the spectral data, $\varepsilon^{\rm tot}_j \cdot k$ governs the normalization for the mass $j$ that represents the injected energy in proportion of the cosmic star formation rate density, $R_{\mathrm{cut}}$ is a cutoff rigidity, $\gamma$ is the spectral index at escape from the sources, and $f_{\rm{cut}} (E_{i}, Z \cdot R_{\mathrm{cut}})=\exp{(1-E_i/(ZR_{\rm cut}))}$ for $E_i>ZR_{\rm cut}$ (1 otherwise) is the cutoff function.

To connect properties at the sources with observables at Earth, we simulate the extragalactic propagation of UHECRs using two well-known Monte Carlo codes in the UHECR community: \texttt{SimProp} \citep{SimProp} and \texttt{CRPropa} \citep{CRPropa}. All generated events are stored in a normalized 5D propagation tensor $T\left(E_{\rm det}, A_{\rm det} \middle| E_{\rm inj}, A_{\rm inj}, z\right)$, which allows us to calculate the probability of detecting a particle injected with energy $E_{\rm inj}$ and mass $A_{\rm inj}$ at a redshift $z$ with an energy $E_{\rm det}$ and mass $A_{\rm det}$. By accounting for possible interactions of UHECRs with the CMB and the different EBL components during propagation, we can calculate the flux at Earth and compare it to observational data. We use the energy spectrum measured in energy bins of 0.1-dex width, ranging from $\log_{10}(E/\mathrm{eV})= 17.8$ to $20.2$, based on the 15-year dataset collected with the surface detector array of the Pierre Auger Observatory \citep{PierreAuger:2021hun}.

Following \citet{Luce_2022} and \citet{Marafico_2024}, we use the deviance
$D = -2\ln\left({\mathcal{L}}/{\mathcal{L}_{\text{sat}}}\right)$,
as an estimator of the agreement between our parametric model and the data, where $\mathcal{L}$ denotes the likelihood of the model and  $\mathcal{L}_{\text{sat}}$ corresponds to a model that perfectly reproduces the  observations. The total deviance is the sum of two contributions, $D_{J}$ and $D_{X_{\rm max}}$. The first term, $D_{J}$, refers to the  energy spectrum and is constructed from Gaussian likelihoods, while the second, $D_{X_{\rm max}}$, is constructed from the comparison of the first two moments (mean and variance) of the $X_{\rm max}$ distributions. In this sense, it is equivalent to a $\chi^2$-type statistic under the assumption of Gaussian uncertainties. In this analysis, we adopt Sibyll2.3d \citep{Riehn_2020} as the hadronic interaction model.

\subsection{Uncertainties on the intrinsic UHECR spectra}

\begin{figure}[!t]
    \centering
    \includegraphics[width=\columnwidth]{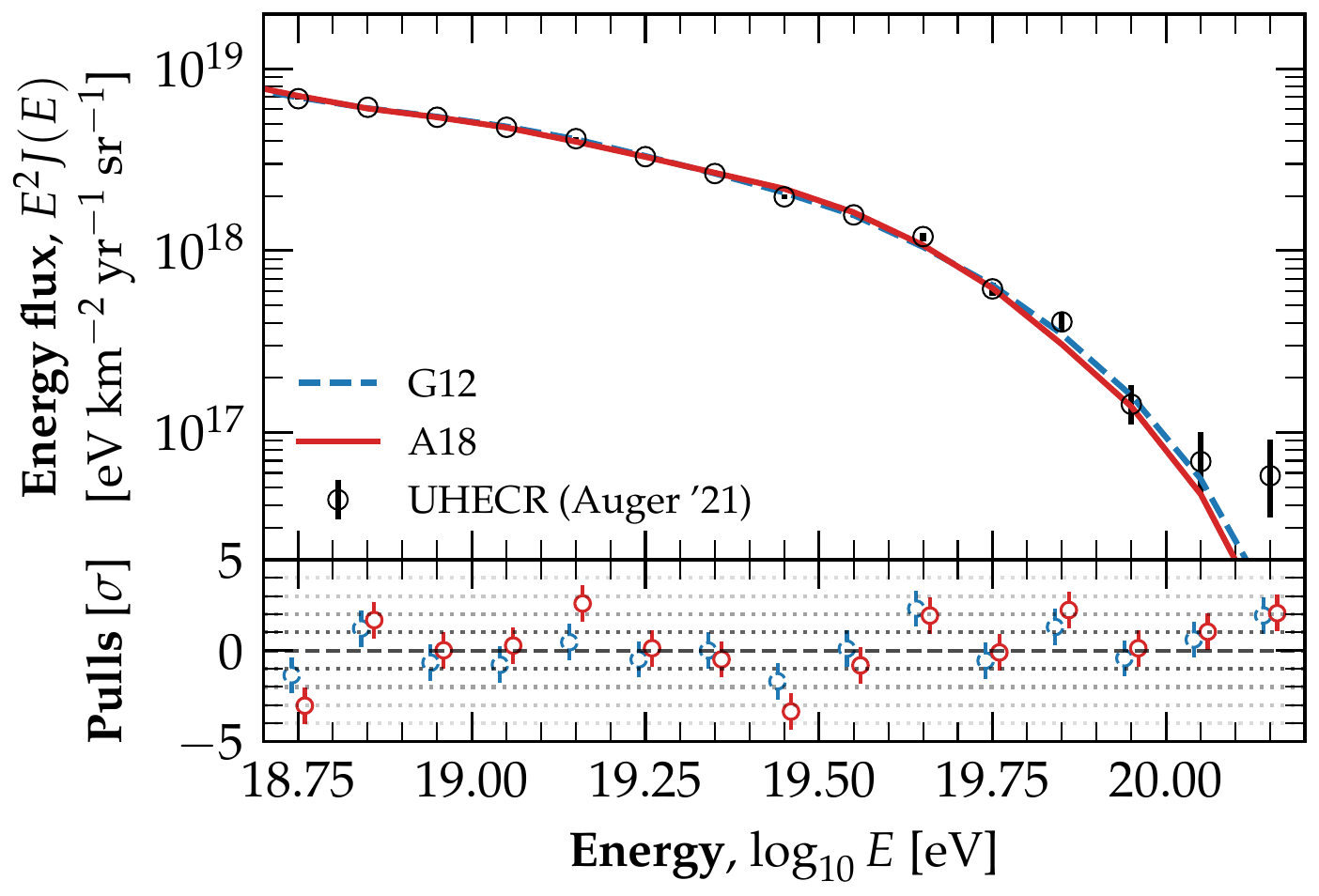}
    \caption{
        \textit{Top}: UHECR all-particle spectrum at Earth. The black points correspond to the measurements of the Pierre Auger Observatory \citep{PierreAuger:2021hun}, and the curves to the all-particle spectra reconstructed using the propagation code \texttt{SimProp} with the specified EBL model.
        \textit{Bottom}: Normalized difference between the UHECR data and the reconstructed spectra. The points have been offset along the $x$-axis for clarity. 
    }
    \label{fig:comparison_data}
\end{figure}

\begin{figure}
    \centerline{\includegraphics[width=1.0\linewidth]{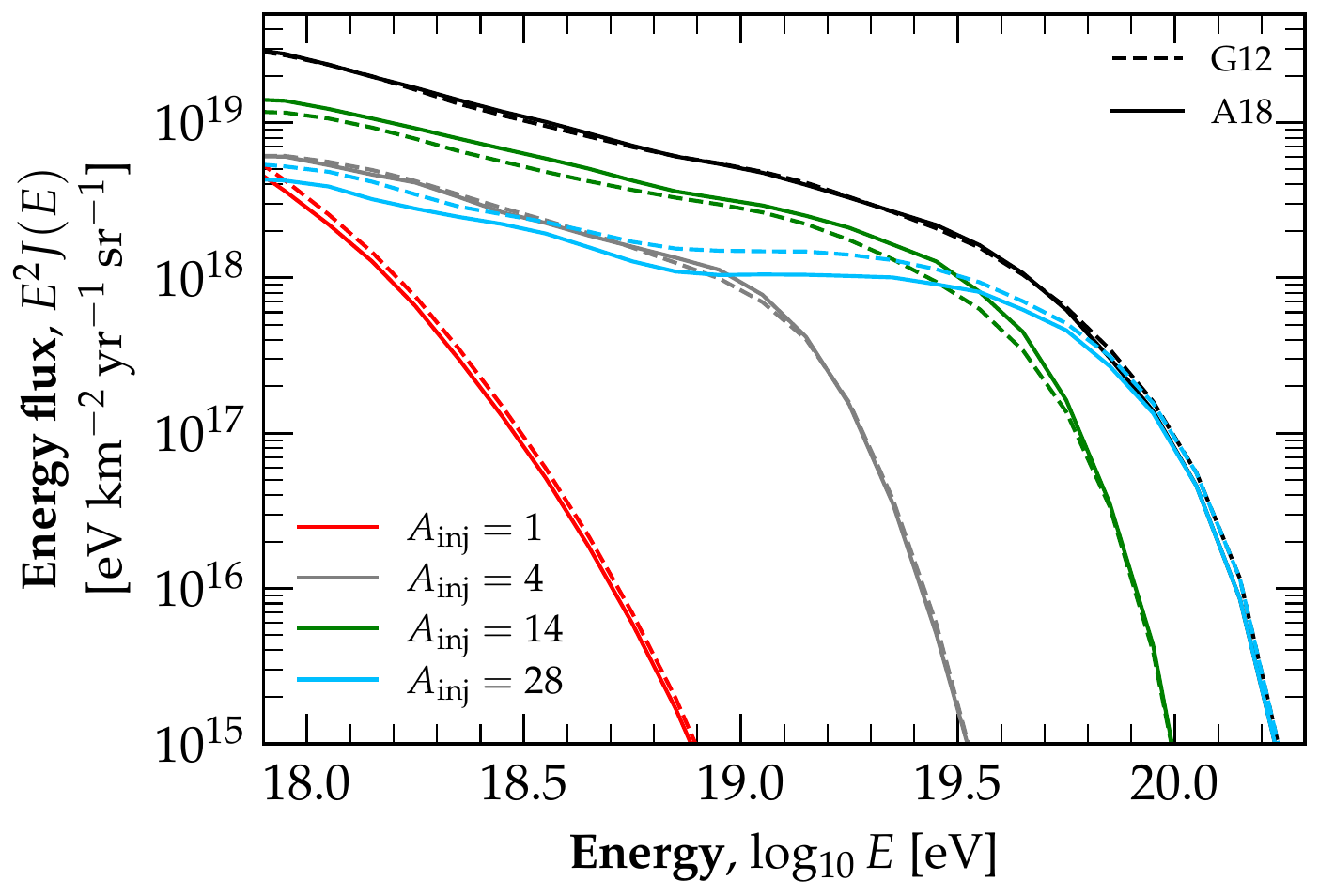}}
    \caption{
        Reconstruction of the contributions of different nuclei injected at the source to the observed spectrum using the \texttt{SimProp} propagation code. Each color corresponds to an injected mass group: proton in red, helium in grey, nitrogen in green and silicon in blue. Iron fluxes are not reported, as they are found to have a negligible overall contribution to the all-particle flux. The black curves correspond to the reconstruction of the flux measured at Earth, as shown in Figure\,\ref{fig:comparison_data}.
    }
    \label{fig:injection}
\end{figure}

To study the impact of the different EBL models and propagation codes on the intrinsic UHECR spectra, we perform a combined fit of the energy spectrum and mass composition of UHECRs for each model and code. Figure\,\ref{fig:comparison_data} shows the best-fit fluxes at Earth obtained for the models G12 and A18, using the \texttt{SimProp} framework. We compare these computed spectra to the UHECR measurements from the Pierre Auger Observatory \citep{PierreAuger:2021hun, AbdulHalim:20239}. The computed spectra at Earth are both compatible with the measurements and with one another. Figure\,\ref{fig:injection} shows the injected spectra at the sources, as reconstructed for both EBL models. There is good agreement between A18 and G12 for all mass groups, with differences of about 20\% across the entire mass range. Similar results are obtained when using the \texttt{CRPropa} propagation code, with differences at the 15\% level.

Previous work by \cite{AlvesBatista:2015jem} studied uncertainties in the propagation of UHECRs arising from differences in the \texttt{CRPropa} and \texttt{SimProp} codes. They found that the uncertainties in the treatment of photodisintegration of $\alpha$-particles were dominant. In \texttt{CRPropa}, all known photodisintegration channels are taken into account.
In \texttt{SimProp}, only two photodisintegration processes are implemented: nucleon and $\alpha$-particle ejection. The interaction rates for these processes are taken to be the sum of all actual processes, weighted by the number of nucleons and $\alpha$-particles ejected, respectively. This ensures the correct total number of free nucleons and those bound in $\alpha$-particles at Earth, though the relative numbers of individual intermediate-mass nuclei may differ.

\begin{figure}[!t]
    \centering
    \includegraphics[width=\columnwidth]{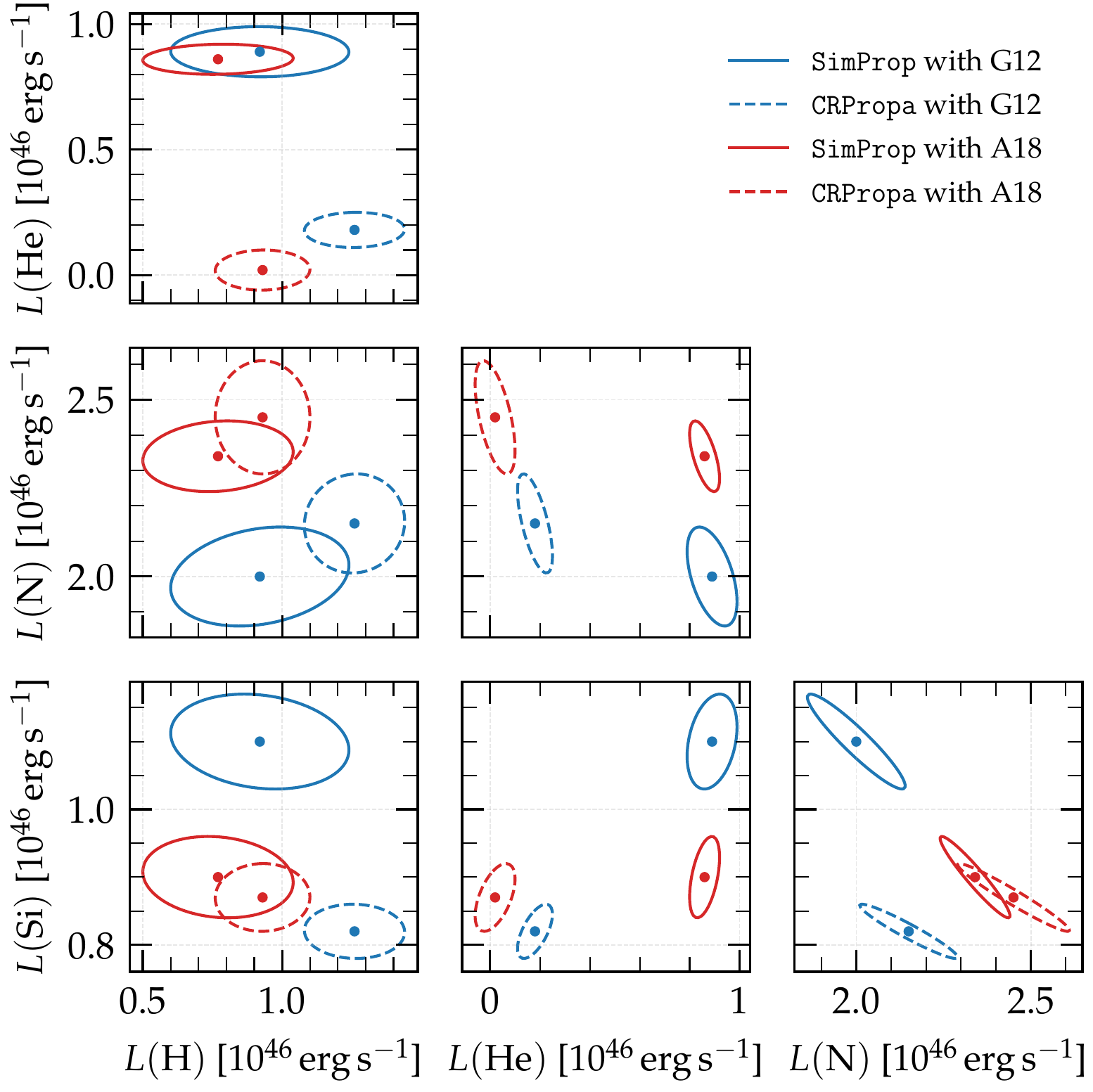}
    \caption{
        Best-fit bolometric luminosities for each mass-component obtained for the different reconstructions.
    }
    \label{fig:corner}
\end{figure}

To compare the impact of each propagation code on the reconstructed UHECR parameters, we retrieved the best-fit luminosities of the proton, helium, nitrogen, and silicon components. We built the parameter vectors $\mmu_\gilmore^\crpropa$, $\mmu_\gilmore^\simprop$, $\mmu_\andrews^\crpropa$, and $\mmu_\andrews^\simprop$, as well as the associated covariance matrices $\boldsymbol{C}_\textrm{model}^\textrm{code}$, and present the pair-wise reconstruction of the parameters in Figure\,\ref{fig:corner}.
In general, we find that the proton and intermediate-mass nucleus luminosities agree within one-to-two standard deviations when changing the EBL model in \texttt{SimProp} and \texttt{CRPropa}, respectively. However, when we replace SimProp with CRPropa for a fixed EBL model, we find that the helium luminosity drops by a factor of $5-10$. This is compensated by a modest $\approx 5$\,\% increase in the more abundant nitrogen luminosity.

\begin{figure}[!t]
    \centering
    \includegraphics[width=\columnwidth]{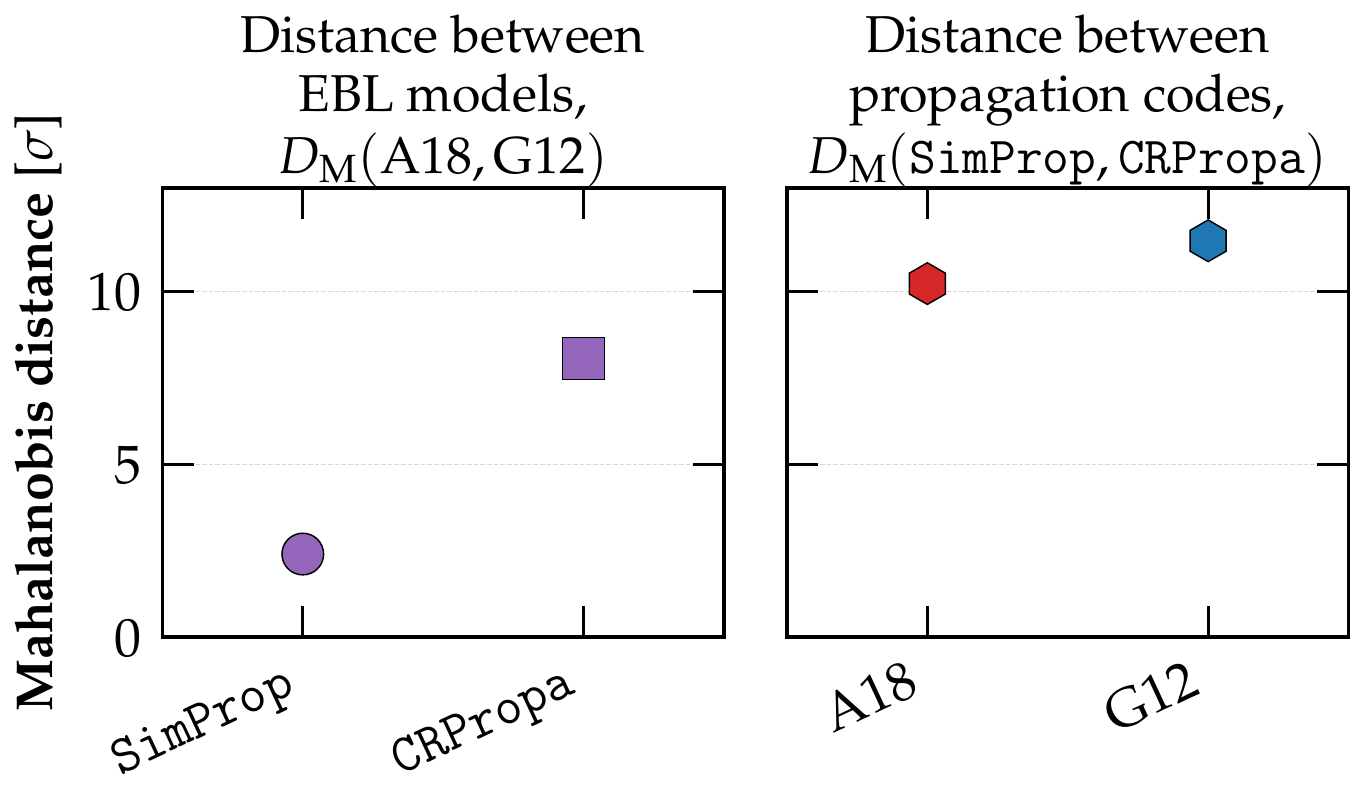}
    \caption{
        Mahalanobis distances between the different reconstructions schemes.
        \textit{Left}: Distances obtained by changing the EBL model used, for a fixed propagation code.
        \textit{Right}: Distances obtained by changing the propagation code used, for a fixed EBL model.
    }
    \label{fig:distances}
\end{figure}

We further quantify the global discrepancy between two reconstruction schemes, $A$ and $B$ (variation of EBL model or variation of the propagation code), using the Mahalanobis distance, defined as follows:
\begin{equation}
    D_{\rm M}(A,B)=\sqrt{(\boldsymbol{\mu}_A-\boldsymbol{\mu}_B)^{\!\top}\,(\mathbf{C}_A+\mathbf{C}_B)^{-1}\,(\boldsymbol{\mu}_A-\boldsymbol{\mu}_B)}\textrm{\,.}
\end{equation}
This distance corresponds to the $l^2$-norm of the difference vector, inverse weighted by the summed covariance. We restrict our comparisons to schemes that share the same choice of either EBL or propagation code, and show the computed distances in Figure\,\ref{fig:distances}. The smallest distance is found between the reconstructions obtained by fixing the propagation code to \texttt{SimProp} using the EBL models G12 and A18. The largest distances are found when using the same EBL model with different propagation codes. As shown in Figure\,\ref{fig:corner}, the proportion of helium injected at the sources is a key factor in explaining why the Mahalanobis distance is larger in the \texttt{SimProp}-\texttt{CRPropa} comparison than in variations of the EBL model.  Depending on the treatment of $\alpha$-particles, the best fit to the data either leads to higher helium injection at the sources with a lower disintegration rate (\texttt{SimProp}), or to higher injection of nitrogen nuclei (\texttt{CRPropa}), which produce $\alpha$-particles along the line of sight — including near Earth.

The impact of the selected EBL models on the bolometric luminosity of each mass component is smaller than the impact of differences between propagation codes used for the reconstruction. However, it should be noted that using EBL models with an infrared component that does not match galaxy counts can result in significant discrepancies from the nominal parameters shown in Figure\,\ref{fig:corner}. For example, the model from \cite{Dominguez_2011}, which is often used as a reference in the UHECR field, yields a Mahalanobis distance of $12$ to $13\sigma$ compared to the A18 or G12 models, which we recommend using instead.

\section{Conclusions} \label{sec:conclusions}

The EBL plays a crucial role in the propagation of both gamma rays and UHECRs. Over the past two decades, the precision of the EBL measurements has steadily increased, particularly due deep-field galaxy surveys. In this study, we leverage this enhanced precision to narrow down the range of EBL models allowed by observations. Our analysis reveals that the models from \citet{Gilmore_2012} and \citet{Andrews_2018} show the greatest compatibility with various measurements of the local EBL intensity. We have explored the impact of the remaining uncertainties on the propagation of gamma rays and UHECRs.

For gamma rays, we find that the choice of EBL model has a negligible impact on the reconstruction of the intrinsic parameters of sources up to a redshift of $z = 0.1$. However, we show that gamma rays from more distant sources enable discrimination between EBL models and, in particular, allows us to probe their evolution with redshift, which direct measurements cannot do.

For UHECRs, we find that uncertainties in EBL models have decreased to the point that they are no longer the primary source of uncertainty. The differences induced by the EBL models selected in this study are below the level of experimental statistical and systematic uncertainties. Additionally, we find that uncertainties of photodisintegration cross sections, and in particular the treatment of $\alpha$ particles, have a larger impact than the choice of EBL model.

In conclusion, our understanding of the cosmic background fields at $z=0$ has advanced to the point that it is no longer the main source of uncertainty in the phenomenological interpretation of UHECR sources and of gamma-ray sources in the low-redshift universe ($z<0.1$). Significant progress is expected in the coming years on multiple fronts. The next generation of Cherenkov telescopes, most notably the Cherenkov Telescope Array Observatory (CTAO), is approaching operation. Additionally, the Pierre Auger Observatory has just finished its AugerPrime upgrade, which aims to improve the reconstruction of the composition of UHECRs. CTAO and Auger will both reduce the statistical and systematic uncertainties in their respective observations, greatly enhancing the ability to detect and characterize populations of VHE gamma-ray sources and the mass composition of the UHECR flux. In parallel, measurements of the EBL spectrum and of its evolution will continue to improve through dedicated observations and future missions, with uncertainties in the IGL expected to reach the percent level~\citep{2021arXiv210212089D}. This convergence will open the door to a new era of high-precision measurements in astroparticle physics.

\subsubsection*{Acknowledgments}
The authors gratefully acknowledge funding from ANR via the grant MultI-messenger probe of Cosmic Ray Origins (MICRO), ANR-20-CE92-0052. This work was made possible by Institut Pascal at Université Paris-Saclay with the support of the program ``Investissements d’avenir" ANR-11-IDEX-0003-01, the P2I axis of the Graduate School of Physics of Université Paris-Saclay, as well as IJCLab, CEA, IAS, OSUPS, and APPEC.

\bibliography{main}{}
\bibliographystyle{aasjournal}

\end{document}